\documentclass[aps,prl,reprint,superscriptaddress,twocolumn,showpacs]{revtex4-1}
\setcitestyle{numbers,square}

\usepackage[T1]{fontenc}
\usepackage[utf8]{inputenc}
\usepackage{textcomp}
\usepackage{amsmath}
\usepackage{mathtools,amssymb,booktabs}

\usepackage{gensymb}

\usepackage{hyperref}
\hypersetup{
  colorlinks=true,
  linkcolor	=black,
  citecolor	=blue,
  urlcolor	=blue
}

\newcommand{\muB}{\mu_\mathrm{B}}

\newcommand{\Tc}{T_\mathrm{c}}

\newcommand{\xx}{\tilde{x}^\text{O}_\text{pl}}

\usepackage{xcolor} 

\begin{document}

\title{Second Dome of Superconductivity in YBa$_2$Cu$_3$O$_7$ at High Pressure}

\author{Johannes Nokelainen}
\email{j.nokelainen@northeastern.edu}
\affiliation{Department of Physics, Northeastern University, Boston, Massachusetts 02115, USA}
\affiliation{Department of Physics and Astrophysics, Howard University, Washington, D.C 20059, USA}
\affiliation{Department of Physics, School of Engineering Science, LUT University, FI-53850 Lappeenranta, Finland}

\author{Matthew E. Matzelle}
\affiliation{Department of Physics, Northeastern University, Boston, Massachusetts 02115, USA}

\author{Christopher Lane}
\affiliation{Theoretical Division, Los Alamos National Laboratory, Los Alamos, New Mexico 87545, USA}

\author{Nabil Atlam}
\affiliation{Department of Physics, Northeastern University, Boston, Massachusetts 02115, USA}

\author{Ruiqi Zhang}
\affiliation{Department of Physics and Engineering Physics, Tulane University, Louisiana 70118 New Orleans, USA}

\author{Robert S. Markiewicz}
\affiliation{Department of Physics, Northeastern University, Boston, Massachusetts 02115, USA}

\author{Bernardo Barbiellini}
\affiliation{Department of Physics, School of Engineering Science, LUT University, FI-53850 Lappeenranta, Finland}
\affiliation{Department of Physics, Northeastern University, Boston, Massachusetts 02115, USA}

\author{Jianwei Sun}
\affiliation{Department of Physics and Engineering Physics, Tulane University, Louisiana 70118 New Orleans, USA}

\author{Arun Bansil}
\email{ar.bansil@northeastern.edu}
\affiliation{Department of Physics, Northeastern University, Boston, Massachusetts 02115, USA}

\date{\today}

\begin{abstract}
Evidence is growing that a second dome of high-$\Tc$ superconductivity 
can be accessed in the cuprates by increasing the doping beyond the 
first dome. 
Here we use \emph{ab initio} methods without invoking any free 
parameters, 
such as the Hubbard $U$, 
to reveal that pressure could turn YBa$_2$Cu$_3$O$_7$ into an ideal 
candidate for second-dome-superconductivity, 
displaying the predicted signature of strongly hybridized $d_{x^2-y^2}$ 
and $d_{z^2}$ orbitals. 
Notably, 
pressure is found to induce a phase transition replacing the 
antiferromagnetic phases with an orbitally-degenerate $d$--$d$ phase. 
Our study suggests that the origin of the second dome is correlated 
with the oxygen-hole fraction in the CuO$_2$ planes and the 
collapse of the pseudogap phase. 

\end{abstract}

\maketitle

In the search for superconductivity (SC) at higher temperatures, 
one of the more intriguing possibilities is that the high-$\Tc$ cuprate 
superconductors also feature another superconducting dome in the 
extremely overdoped metallic and non-magnetic (NM) 
regime~\cite{2006_Uchida_SCO_95K_Tc,
             2009_Geballe_SCO_Tc, 
             2010_Chen_Bi2223_pressure_two_SC_domes,
             2016_Zhong_CuO2Bi2212_nodeless_pairing_exp,
             2018_Jiang_CuO2+Bi2212_model_DFT,
             2019_Oles_high-Tc_orbital_excitations_review, 
             2019_Li_PNAS_BCO_discovery,
             2019_Scalapino_PRB_2nd_dome_Cu3+,
             2019_Scalapino_PNAS_different_branch_high-Tc,
             2019_Chu_2nd_dome_pressure_bisco,
             2020_Conradson_PNAS_SrCuO_structure,
             2020_Matsumoto_bilayer_hubbard_model_2nd_dome,
             2021_Sederholm_high-P-oxygenation_cuprate_review,
             2021_Zegrodnik_CuO2+Bi2212_model}. 
This is in stark contrast to the long-studied `first dome' that can be 
characterized as a lightly doped Mott insulator with a partially 
occupied Cu $d_{x^2-y^2}$ band and short-range antiferromagnetic (AFM) 
order~\cite{1989_Pickett_cuprate_megareview}. 
Evidence for the existence of such a `second dome' has been building in 
recent years, 
as SC has been observed even when the doping levels have been driven 
far beyond the first dome via high-pressure oxygenation (HPO) 
synthesis~\cite{2021_Sederholm_high-P-oxygenation_cuprate_review}. 
Findings of 
$\Tc = 95$\,K in Sr$_2$CuO$_{4-y}$~\cite{2006_Uchida_SCO_95K_Tc} and 
$\Tc = 70$\,K in 
Ba$_2$CuO$_{4-y}$~\cite{2019_Li_PNAS_BCO_discovery} 
have heightened interest since these $\Tc$s are clearly higher than 
$\Tc \sim 44$\,K of their isostructural counterpart 
La$_2$CuO$_{4+\delta}$. 
Also, 
a variety of other types of unconventional superconductors have been 
shown to feature two or more SC domes with distinct 
characteristics~{\cite{2016_Das_2domes_summary,
                      2018_Shimizu_FeS_two_SC_domes,
                      2020_Reiss_2_domes_iron_superconductor,
                      2021_Feng_kagome_two_SC_domes,
                      2021_Chen_kagome_two_SC_domes,
                      2021_Liu_two_domes_multiorbital_superconductors}. 

To explain the second dome, 
Maier \emph{et al.}\ have constructed a model where the valency is 
increased by introducing holes on the Cu $d_{z^2}$ orbitals, 
resulting in a new peak in the pairing 
function~\cite{2019_Scalapino_PRB_2nd_dome_Cu3+, 
               2019_Scalapino_PNAS_different_branch_high-Tc}. 
Similar models~\cite{2018_Jiang_CuO2+Bi2212_model_DFT, 
                     2021_Zegrodnik_CuO2+Bi2212_model} 
have been invoked to explain the presence of a second dome in a single 
CuO$_2$ plane grown on top of 
Bi$_2$Sr$_2$Ca$_{n-1}$Cu$_n$O$_{2n+4+\delta}$ (BSCCO), $n=2$~\cite{2016_Zhong_CuO2Bi2212_nodeless_pairing_exp}. 
However, 
this system is not practical for transport measurements, 
while for the HPO-cuprates the polycrystalline samples contain many 
SC phases with a multiplicity of oxygen vacancy orderings and it is 
unclear which of the phases are the best 
superconductors~\cite{2020_Conradson_PNAS_SrCuO_structure}. 

Besides chemical doping, 
pressure ($P$) is also known to increase doping 
levels~\cite{2022_Mark_cuprate_pressure_review} 
and 
enhance $T_c$ in a number of cuprates~\cite{schilling2007high}. 
Pressure has also been observed to trigger a second, 
anomalous, 
rise of $\Tc$ in BSCCO, 
$n=1,2,3$; 
when the pressure is applied to optimally doped samples, 
$\Tc$ first decreases but then increases into a second 
dome~\cite{2010_Chen_Bi2223_pressure_two_SC_domes,
           2019_Chu_2nd_dome_pressure_bisco}. 
However, 
the notoriously complicated multi-phase structure of BSCCO involving 
supermodulation and complex oxygen 
ordering~\cite{2011_Poccia_Bi2212_supermodulation_X-ray,
               2014_Zeljkovic_bisco_Oint_periodicity,
               2019_Song_2212_Oint+Superstructure_STM_DFT} 
makes it difficult to characterize the underlying physics and the 
second dome has not been observed in other recent 
studies~\cite{2021_Matsumoto_bisco_whiskers_pressure,
              2022_Zhou_Bi2212_insulator_under_pressure}. 

Here, 
we present an {\it ab initio} study of pressure-induced 
doping of the prototypical cuprate superconductor YBa$_2$Cu$_3$O$_7$ 
(YBCO$_7$) and discuss its high-pressure states from the viewpoint of 
the second dome physics. 
YBCO$_7$ offers the advantages that it is in the overdoped regime 
yet stoichiometric, 
as illustrated in Fig.~\ref{fig1}\,(a). 
Thus, 
pressure can more easily drive the material to the extremely overdoped 
regime in which the structure remains single-phase, 
reducing the challenges associated with complex 
distributions and pressure effects of dopant atoms.
Our first-principles computations employ the 
strongly-constrained-and-appropriately-normed (SCAN) 
density-functional~\cite{Blochl1994_PAW,
                         Kresse1999_PAW,
                         1996_Kresse_VASP_PRB,
                         2015_Sun_SCAN}, 
which has been shown to provide a good parameter-free first-principles 
description of electronic correlation and Cu--O charge transfer 
physics~\cite{2018_Chris_La2CuO4_SCAN, 
               2018_Furness_SCAN_cuprate, 
               2020_Nokelainen_bisco,
               2020_Yubo_SCAN_stripe_YBCO,
               2022_Tatan_Hg-cuprates_SCAN,
               2022_Kanun_functional_comparisons,
               2023_Ruiqi_nickelate_nemacity,
               2023_Yale_cuprate_normal_state_first_principles,
               2024_Jinliang_YBCO_phonons_comparisons}. 
Various magnetic orders are considered under hydrostatic pressure, 
which we increase adiabatically from zero up to 170\,GPa. 
The Supplemental Material (SM)~\cite{SM} contains the details of our 
computations. 
Our analysis shows that pressure induces three key effects in YBCO$_7$ 
that facilitate the transition to the second dome: 

\begin{figure}[htpb]
\centering
\includegraphics[width=\linewidth]{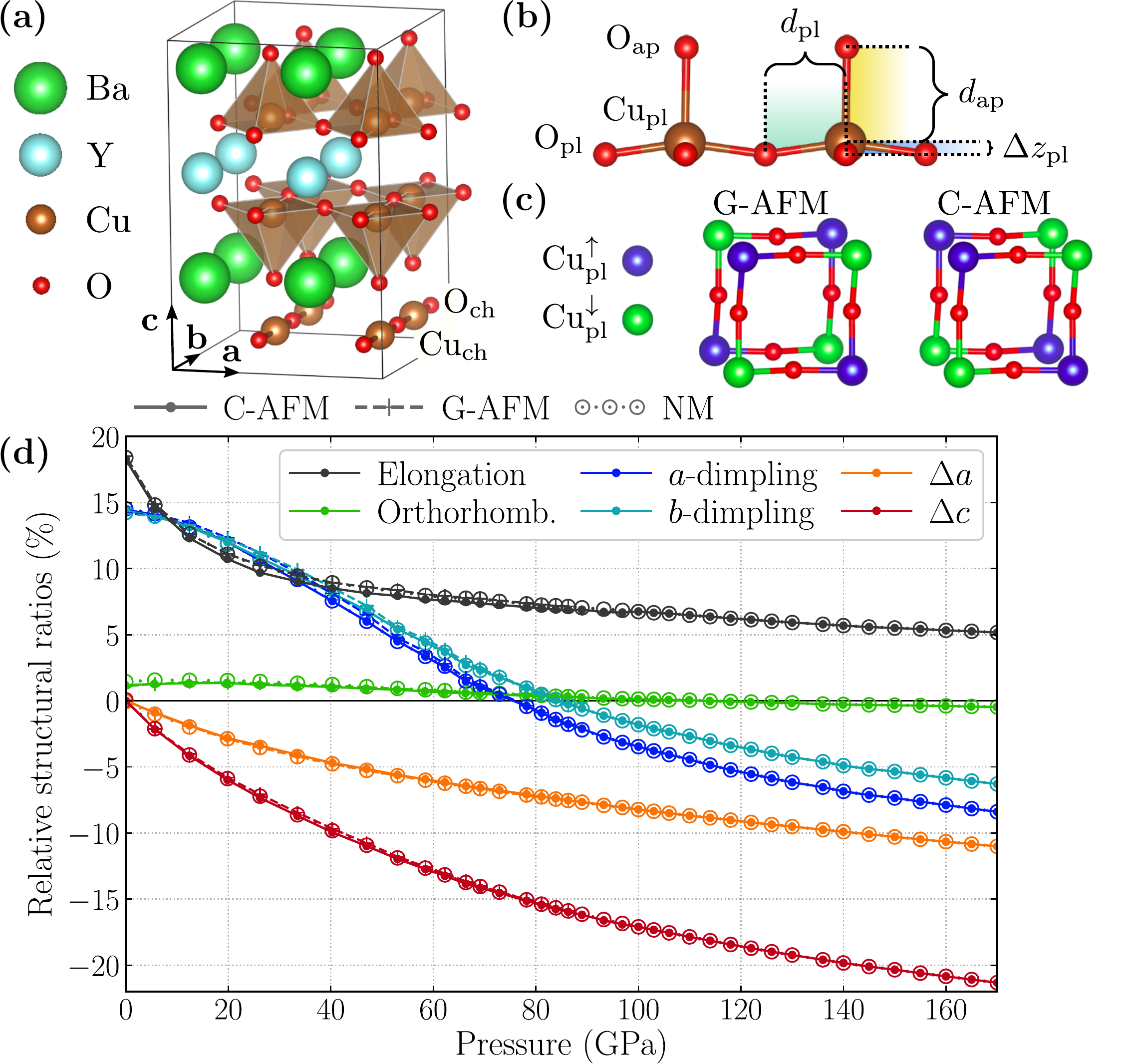}
\caption{(a): 
 Atomic structure of YBCO$_7$. 
 Cu$_\text{pl}$ atoms reside in pyramidal cages formed by the 
 O$_\text{pl}$ and O$_\text{ap}$ atoms. 
 The CuO$_2$ planes are separated by a Y buffer layer. 
 Sandwiched between the BaO layers are the CuO chains, 
 which contain the oxygen dopants. 
 Note that YBCO$_7$ is the fully doped compound of the 
 YBCO$_{6+\delta}$ family with complete CuO chains, 
 while YBCO$_6$ is the undoped system in which all the 
 O$_\text{ch}$ sites are vacant. 
 The chains are arranged along the $b$-axis, 
 which is 1.2\,\% longer than the $a$-axis ($a_0 = 3.83\,$\AA). 
 (b): Definitions of the $d_\text{pl}$ and $d_\text{ap}$ Cu--O 
 distances and $\Delta z_\text{pl}$. 
 (c): Schematics of the G-AFM and C-AFM magnetic configurations.
 (d): In \% units, 
 $P$-evolution of relative elongation of the Cu--O pyramids 
 $(d_\text{ap} - d_\text{pl})/d_\text{pl}$,  
 relative orthorhombicity of the unit cell $(b-a)/a$, 
 dimpling ratio $\Delta z_\text{pl}/d_\text{pl}$ 
 (separately for the $a$- and $b$-directions) 
 and lattice parameter modifications 
 $\Delta a = (a-a_0)/a$ and $\Delta c = (c-c_0)/c$. 
 \label{fig1}
}
\end{figure}

\emph{(1): Change in the pyramidal Cu--O environment}. 
Figure~\ref{fig1}\,(d) presents the pressure-evolution of several 
structural parameters. 
There are no substantial structural differences between the NM state 
and the G-AFM and C-AFM orders, 
see Fig.~\ref{fig1}\,(c) for definitions. 
The lattice constants are in excellent agreement with 
Ref.~\cite{2009_Calamiotou_YBCO7_lattice_pressure}
(within 0.4\,\% at $P=0$ and 0.7\,\% at $P\approx12.7$\,GPa). 
However, 
we did not observe the structural instabilities found in 
Ref.~\cite{2009_Calamiotou_pressure_instablities}, 
see SM Sec.~S6~\cite{SM} for details. 
As is typical of layered materials such as the cuprates, 
the $c$-axis shrinks more rapidly than the $a$ and $b$-axes 
(e.g.\ at 140\,GPa $c$ has decreased by about 20\,\% 
while $a$ and $b$ have decreased by about 10\,\%). 
Concurrently, 
there is a decrease in the elongation of the pyramidal Cu--O cages, 
which is intimately connected to the relative energies of the 
Cu-$d_{x^2-y^2}$ and $d_{z^2}$ orbitals. 
Interestingly, 
enhanced elongation is known to be favorable for the first dome of 
SC~\cite{2012_Sakakibara_cuprate_pressure_models_DFT} 
while the opposite seems to be the case in 
Ba$_2$CuO$_{4-y}$~\cite{2019_Li_PNAS_BCO_discovery}. 
YBCO$_7$ has a relatively large elongation of about 18\,\% 
but it is quickly reduced under pressure and at 170\,GPa it is only 
about 5\,\%. 
Notably, 
the strong dimpling [Fig.~\ref{fig1}\,(b)] 
of $14.2\,\%$ is reduced under pressure until it reverses sign at 
80\,GPa such that the Cu$_\text{pl}$ atoms bulge out from the pyramids 
towards the Y layer. 
At $170\,$GPa the dimpling is $-8.4\,\%$ ($-6.3\,\%$) in the $a$ 
($b$)-direction. 
Also the slight non-orthorhombicity $(b>a)$ of 2\,\% is removed under 
pressure and weakly reversed above 120\,GPa. 

\emph{(2): Pressure-induced doping}. 
Figure~\ref{fig2}\,(a) presents the hole contents ($x$) based on the 
Bader charge analysis~\cite{2009_Tang_Bader,
                            2011_Yu_Trinkle_Bader_improvement}. 
The zero-pressure doping on the CuO$_2$ planes is 
$x([\text{CuO$_2$}]_\text{pl}) = 
x(\text{Cu}_\text{pl}) + 2 x(\text{O}_\text{pl}) = 0.11$ 
holes~\footnote{The value $x([\text{CuO$_2$}]_\text{pl})=0.11$ 
  holes is smaller than the value 0.18 holes found in 
  the literature for YBCO$_7$~\cite{2006_Liang_YBCO_doping_formula}; 
  this difference is discussed in the SM Sec.~S3.1~\cite{SM}.} 
and the CuO chains assume a negative doping of 
$x([\text{CuO}]_\text{ch}) = -0.57$ holes, 
with weak magnetic configuration dependence. 
This reflects charge transfer from the CuO$_2$ planes to the CuO 
chains as YBCO$_6$ is doped to YBCO$_7$. 
Significant doping ($0.11$ holes) is found also on the O$_\text{ap}$ 
sites. 
Under pressure, 
the CuO$_2$ plane is doped further 
up to $0.28$ holes at 170\,GPa, 
an increase of $0.17$ holes over the zero-pressure value. 
The sources of this doping are the Ba and Y ions, 
which capture electrons at an almost linear rate until about 60\,GPa, 
after which this electron capture rate slows down, 
especially for Ba. 
This could simply be the result of the large ionic radii of Y and Ba.
When the lattice contracts under pressure, 
$e$--$e$ repulsion would tend to force the electrons to migrate from 
the tightly packed CuO$_2$ planes to the buffer layers with smaller 
electron density. 
Interestingly, 
the $P$-induced doping has been previously attributed only to the CuO 
chains, 
either due to an increased charge transfer between the chains and the 
CuO$_2$ planes or to the reordering of the O$_\text{ch}$ 
vacancies~\cite{1990_Jorgensen_YBCO7_pressure_exp,
                1992_Almasan_YBCO7_pressure_exp,
                2000_Sadewasser_YBCO_pressure_chain_ordering,
                2018_Cyr-Choiniere_YBCO7_pressure_CDW_suppression}. 
However, 
we find that the chains also become doped with $0.09$ holes 
over the pressure range considered. 

\begin{figure}[htpb]
\includegraphics[width=\linewidth]{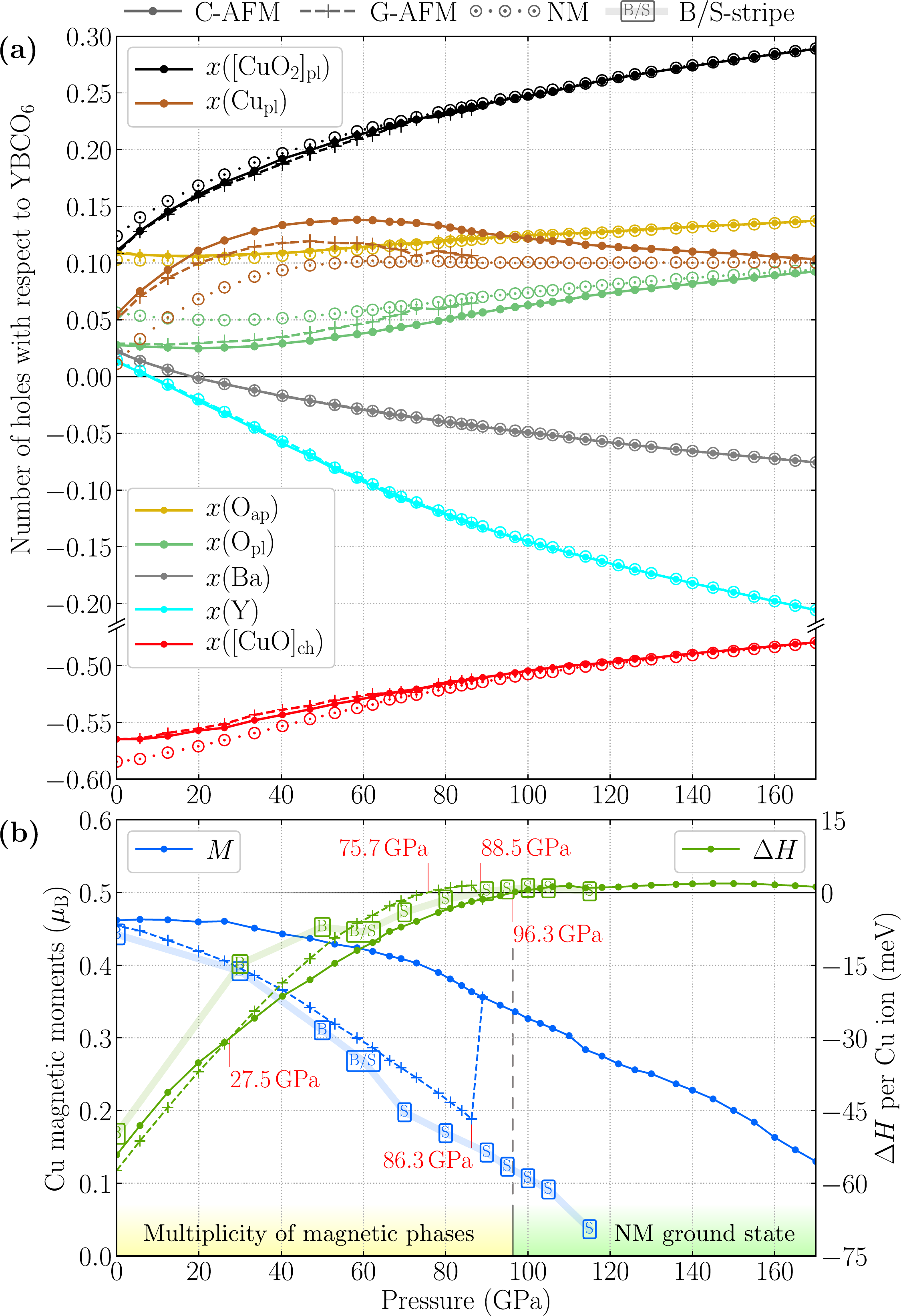}
\caption{(a): Hole contents for various ions and groups obtained 
  as Bader charge differences between YBCO$_7$ and the undoped base 
  compound YBCO$_6$, 
  see SM Sec.~S3~\cite{SM} for details. 
  (b): \emph{Left axis:} 
  Pressure-evolution of the Cu$_\text{pl}$ magnetic moments. 
  \emph{Right axis:} Magnetic enthalpy 
  $\Delta H = H_\text{AFM} - H_\text{NM}$. 
}\label{fig2}
\end{figure}

The distribution of the doping within the CuO$_2$ planes deserves 
special attention. 
Below $P\approx 40\,$GPa, 
$x(\text{Cu}_\text{pl})$ grows rapidly while $x(\text{O}_\text{pl})$ 
stagnates, 
i.e., 
the additional doping ends up on the Cu ions. 
But this trend starts to change at higher pressures. 
Beyond 80\,GPa the situation is reversed; 
$x(\text{Cu}_\text{pl})$ even starts to decrease 
while $x(\text{O}_\text{pl})$ grows steadily. 
We will return to discuss the implications of this behavior below. 

\emph{(3): Stabilization of the NM state}. 
Figure~\ref{fig2}\,(b) presents the Cu$_\text{pl}$ magnetic moments ($M$) 
and enthalpies relative to the NM phase ($\Delta H$) 
for the studied magnetic phases. 
At zero pressure the G-AFM and C-AFM states 
and a $3\times2$ bond-centered stripe (B-stripe) phase~\footnote{
  This stripe phase was the most stable $2\times3$ or $3\times2$ 
  stripe phase in our pressure tests, 
  see SM Sec.~S4~\cite{SM} for details. 
  Larger stripe phases could be energetically more favorable 
  at zero pressure~\cite{2020_Yubo_SCAN_stripe_YBCO} 
  but considering them was not feasible due to computational reasons. 
} 
are below the NM state by 57\,meV, 54\,meV and 50\,meV,
respectively, 
which is consistent with the multiplicity of near degenerate 
magnetic phases found in Ref.~\cite{2020_Yubo_SCAN_stripe_YBCO}. 
A direct effect of the pressure-induced doping is the suppression of 
the magnetic moments. 
$M_\text{G-AFM}$ and $M_\text{stripe}$ 
start to decrease as soon as pressure is applied, 
but $M_\text{C-AFM}$ is robust up to 30\,GPa 
before it starts to decrease. 
The pressure also decreases the stability of the magnetic phases, 
but the competition between them persists. 
The C-AFM state becomes the new ground state over the G-AFM 
state at 27.5\,GPa. 
The G-AFM state further rises above the NM phase at 75.7\,GPa 
and transitions to the C-AFM state at pressures beyond 86.3\,GPa. 
The B-stripe also goes through a phase transition into a 
site-centered stripe phase (S-stripe) around 60\,GPa 
(where it shows mixed characteristics) 
and rises above the NM phase at 88.5\,GPa. 
Finally, 
the C-AFM state becomes metastable at 96.3\,GPa 
so that the NM state is the ground state at high pressures. 
In this NM region, 
we were able to preserve the C-AFM phase 
up to the highest studied pressure of 170\,GPa 
and S-stripe up to 115\,GPa 
by increasing pressure adiabatically~\footnote{At these 
  pressures these states become fragile and can be preserved 
  only with the most accurate technical settings, 
  see SM Sec.S1.4~\cite{SM} for details.}. 
Notably, 
the C-AFM state remains robust about 
$1.5\,$meV above the NM state while the magnetic moments weaken. 
At 170\,GPa, 
$M_\text{C-AFM}=0.13\,\muB$, 
but we expect the C-AFM phase to finally vanish at still higher 
pressures. 
The relative stability of the C-AFM configuration could be due to its 
ferromagnetic interlayer coupling [Fig.~\ref{fig1}\,(c)], 
which allows covalent bonding between the magnetic Cu orbitals when the 
pressure moves the CuO$_2$ planes closer to each other. 

\begin{figure}[htpb]
\centering
\includegraphics[width=\linewidth]{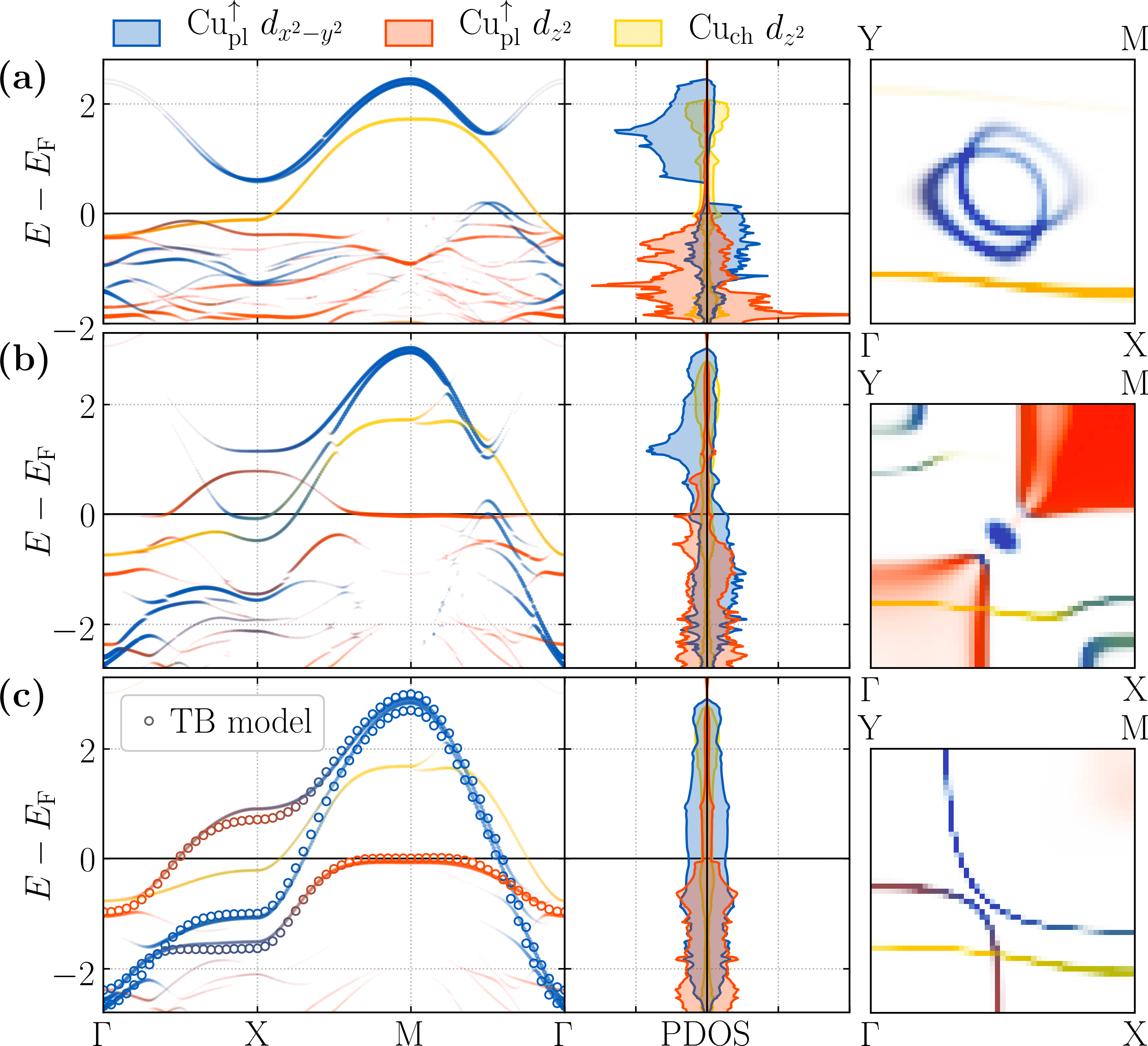}
\caption{Orbital-decomposed YBCO$_7$ electronic structures and 
  Fermi surfaces (spectral functions at the Fermi energy) 
  at $k_z=0$ for Cu$_\text{pl}$ $d_{x^2-y^2}$, 
  Cu$_\text{pl}$ $d_{z^2}$ and Cu$_\text{ch}$ $d_{z^2}$ for 
  (a) $P=0$ G-AFM phase ($M=0.462\,\muB$), 
  (b) $100$\,GPa C-AFM phase ($M=0.326\,\muB$) and 
  (c) $100$\,GPa NM phase. 
  In the AFM cases the Cu$_\text{pl}$ PDOS has been plotted 
  only for ions with positive $M$ and the energy bands and 
  spectral functions have been unfolded into the primitive NM cell. 
  The three highest bands of our four-band TB model are plotted for 
  the $100$\,GPa NM case. 
}\label{fig3}
\end{figure}

\emph{Electronic structure under pressure.} 
Figure~\ref{fig3}\,(a) shows the YBCO$_7$ electronic structure at 
ambient pressure for the G-AFM state, 
which is typical of the cuprates---dominated by 
planar Cu $d_{x^2-y^2}$. 
Also chain Cu $d_{z^2}$ bands are present, 
but they only have negligible hybridization with the 
Cu$_\text{pl}$ $d_{x^2-y^2}$ bands. 
The bilayer splitting leads to two distinct $d_{x^2-y^2}$ Fermi surface 
rings that partly disappear due to unfolding into the primitive NM unit 
cell~\cite{2010_Wei_band_unfolding, pyprocar}. 
The $d_{z^2}$ states display Hund's coupling with the $d_{x^2-y^2}$ 
bands~\cite{2018_Chris_La2CuO4_SCAN}, 
as seen from the spin polarization in the partial density of states 
(PDOS), 
and have a slight Fermi level contribution through hybridization with 
the $d_{x^2-y^2}$ and chain bands. 

At 100\,GPa, 
the $d_{z^2}$ band rises to the Fermi level for both the metastable 
C-AFM state [Fig~\ref{fig3}\,(b)] and the NM ground state 
[Fig.~\ref{fig3}\,(c)]. 
This increases the Cu$_{\text{pl}}$ $d_{z^2}$ hole content 
at 100\,GPa despite of decrease in the total Cu$_{\text{pl}}$ hole 
content, 
see SM Sec.~S3.3~\cite{SM} for details. 
The 100\,GPa C-AFM state has little hybridization between the two $d$ 
orbitals but significant $d_{x^2-y^2}$--chain hybridization. 
The $d_{z^2}$ bands are pinned to the Fermi level, 
possibly because they are pushed upwards by Hund's coupling but acquire 
holes at a slow rate. 
This flat band dominates the states at the Fermi energy ($k_z=0$), 
but due to slight three-dimensionality of the $d_{z^2}$ bands this is less 
prominent for $k_z \neq 0$, 
see SM Sec.~S2.3~\cite{SM} for details. 
Once the magnetization is suppressed, 
in the NM phase, 
$d_{z^2}$ and $d_{x^2-y^2}$ hybridize strongly around the $X$ point 
(where the AFM bands overlap) 
and the $d_{x^2-y^2}$--chain hybridization is absent, 
contrary to the C-AFM case. 
The NM phase spectral function at Fermi energy has 
mainly $d_{x^2-y^2}$ and chain contributions and the $d_{z^2}$ flat 
band is only weakly visible around $M$. 

To explore the second dome physics under pressure, 
we have adopted a minimal NM tight binding (TB) model that accounts for 
bilayer splitting and $d_{x^2-y^2}$---$d_{z^2}$ 
hybridization~\cite{2014_Baublitz_manganite_minimal_TB_dx2_dz2}. 
The model is overlaid in frame (c) of Fig.~\ref{fig3} for $P=100$\,GPa 
and plotted for $P=0$ in the SM Fig.~S10~\cite{SM}. 
The model is based on using symmetric and anti-symmetric intra-orbital 
combinations, 
so that the Hamiltonian becomes block-diagonal. 
The strong bilayer splitting between the $d_{z^2}$ orbitals shifts the 
anti-symmetric $d_{z^2}$ band to much lower energies. 
This separation keeps the anti-symmetric $d_{z^2}$ and $d_{x^2-y^2}$ 
bands from hybridizing. 
However, 
the symmetric bands remain close in energy, 
leading to the strong hybridization near the $X$-point. 
Our symmetric bands are strikingly similar to Maier~\emph{et al.}'s 
model at $x=0.85$~\cite{2019_Scalapino_PRB_2nd_dome_Cu3+}, 
which is within their second dome. 
The two TB models can be directly mapped onto each other with the 
exception of a few next-nearest neighbor terms, 
see SM Sec.~S7~\cite{SM}. 
Based on this intimate connection, 
NM YBCO$_7$ at 100\,GPa sits within the underdoped side of the second 
dome predicted by Ref.~\cite{2019_Scalapino_PRB_2nd_dome_Cu3+}. 

\begin{figure}[htpb]
\includegraphics[width=\linewidth]{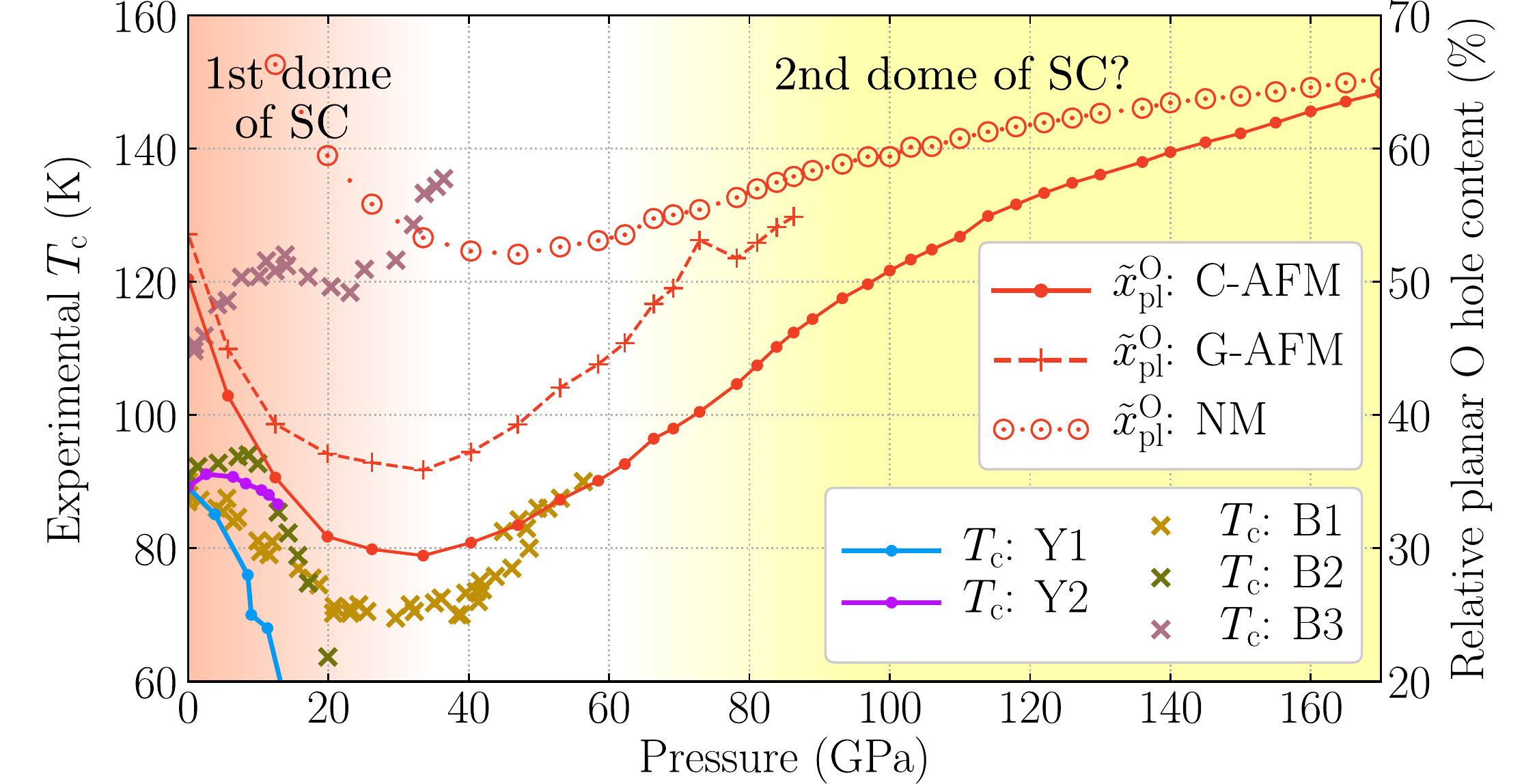}
\caption{\emph{Left axis:} 
  Experimental $P$-dependence of $\Tc$ in selected samples; 
  Y1: YBCO$_{6.98}$~\cite{2017_Tacon_YBCO7_28GPa}. 
  Y2: YBCO$_{6+\delta}$ with $\delta$ estimated to be between 
  0.85 and 1.00~\cite{1991_Klotz_YBCO7_pressure_exp}.
  B1: Slightly overdoped $n=2$ 
  BSCCO~\cite{2019_Chu_2nd_dome_pressure_bisco}. 
  B2 and B3: Optimally doped 
  $n=2$~\cite{2010_Chen_Bi2223_pressure_two_SC_domes} and 
  $n=3$~\cite{2022_Zhou_Bi2212_insulator_under_pressure} BSCCO. 
  \emph{Right axis:} 
  The fractional planar O hole content obtained from the Bader analysis 
  as $\xx = 2x(\text{O}_\text{pl}) / x([$CuO$_2]_\text{pl})$. 
}\label{fig4}
\end{figure}

\emph{Importance of O$_\text{pl}$ hole content:} 
The long-sought explanation for the pairing mechanism in the first dome 
might have recently been verified to be 
superexchange~\cite{2022_Davis_cuprate_superexchange_experiment}, 
as proposed just a few months after the discovery of 
high-$\Tc$s~\cite{1987_Anderson_LCO_superexchange_resonating_VB}. 
There is a correlation between the maximal $\Tc$ and the planar O 
hole content~\cite{2016_Haase_Uemura_generalization}, 
which can be understood within the superexchange 
model~\cite{2021_Tremblay_cuprate_O_hole_content_model}. 
In this spirit, 
we plot in Fig.~\ref{fig4} the fraction of the YBCO$_7$ oxygen holes 
within the CuO$_2$ planes ($\xx$), 
as well as the available experimental $\Tc$ data, 
which we supplement with BSCCO data that extends to higher pressures. 
Below 20\,GPa, 
the pressure-induced doping leads to the suppression of $\Tc$ in YBCO 
(Y1 and Y2) and $n=2$ BSCCO samples (B1 and B2), 
which is consistent with these samples either being close to 
optimal doping 
or sitting on the overdoped side 
of the first SC dome. 
Pressure also causes $\xx$ to decrease, 
which is consistent both with the superexchange model and 
Ref.~\cite{2021_Tremblay_cuprate_O_hole_content_model}, 
in that it shows a correlation between $\Tc$ and $\xx$ for the first 
dome. 
However, 
recent YBCO$_{6.9}$ low-pressure Cu--O charge transfer 
results~\cite{2023_Haase_YBCO_pressure} based on nuclear magnetic 
resonance measurements~\cite{2004_Haase_NMR_cuprate_hole_content} are 
somewhat different from our results, 
see SM Sec.~S3.2~\cite{SM} for a discussion of this point. 

Based on our TB analysis, 
we expect YBCO$_7$ to enter the second dome regime around $100$\,GPa. 
Even though experimental $\Tc$ data for YBCO$_7$ is limited, 
the B1 and B3 BSCCO datasets display a pressure-induced revival in SC 
above $\sim30$\,GPa. 
Also, 
$\xx$ begins to increase with pressure, 
with striking similarity to the B1 dataset. 
At 100\,GPa, 
the $\xx$ reaches the $P=0$ AFM values, 
raising the possibility of a correlation between $\xx$ and $\Tc$ in the 
second dome. 
If so, 
how does this affect the pairing mechanism? 
Here there are many more possibilities than in the case of the first dome 
since both Cu $d_{x^2-y^2}$ and $d_{z^2}$ are involved. 
Notably, 
Ref.~\cite{2019_Scalapino_PRB_2nd_dome_Cu3+} finds two pairing channels:
$d$-wave 
and $s^{\pm}$-wave, 
with both channels dominated by spin fluctuations of the $d_{z^2}$ 
electrons. 
Scenarios for multi-orbital pairing have been discussed in 
Ref.~\cite{2019_Oles_high-Tc_orbital_excitations_review}. 
In the orbital fluctuation model---another early proposal for 
explaining high $\Tc$---pairing between oxygen holes is mediated by the 
$d$-orbitals~\cite{1988_Weber_cuprate_excitation_model, 
                   1988_Jarrell_charge_transfer_mechanisms_for_high-Tc, 
                   1989_Cox_quadrupole_fluctuation_high-Tc_mechanism}, 
potentially explaining the importance of $\xx$ and the oxygen holes 
for the second dome. 
Alternatively, 
the pressure-induced enhancement in $\xx$ and intraplanar covalency 
could revive the superexchange mechanism for the second 
dome~\cite{2021_Tremblay_cuprate_O_hole_content_model} 
or the $d_{z^2}$ flat band could induce pairing-effective 
$s^\pm$ spin 
fluctuations~\cite{2020_Matsumoto_bilayer_hubbard_model_2nd_dome}, 
as discussed in the SM~Sec.~S5.2~\cite{SM}. 
Finally, 
we note that the near degeneracy of the NM state with the metastable 
high-pressure C-AFM state could lead to enhanced pairing fluctuations.

\emph{Relationship with pseudogap collapse:}
Cuprates are characterized by a mysterious pseudogap phase that involves
intertwined orders~\cite{2015_Tranquada_high-Tc_intertwined_orders,
    2019_Schmalian_vestigal_order_in_QM,
    2019_Michon_thermodynamic_signatures_of_quantum_criticality,
    2021_Taillefer_Bi2201_pseudogap,
    2024_markiewicz_arxiv_gapsend_afm_domain_walls}. 
The abundance of near-degenerate stripe phases found in ambient-pressure 
YBCO$_7$ have been proposed to be a signature of the pseudogap 
phase~\cite{2020_Yubo_SCAN_stripe_YBCO}. 
Our results show that the multiplicity of these phases persists to 
high pressures, 
where these phases rise about $1.5\,$meV above the NM state 
(which is 54\,meV above the ground state without pressure), 
indicating termination of the pseudogap phase. 
This is reminiscent of La$_2$CuO$_4$, 
where the NM phase is 150\,meV above the ground state at zero 
doping~\cite{2018_Chris_La2CuO4_SCAN} while the magnetic phases have 
been pushed up to 60\,meV above the NM phase at 
30\% doping~\cite{2007_wakimoto_lsco_pseudogap}. 
We also observe a Van Hove singularity crossing the Fermi level near 
126\,GPa, 
another signature of pseudogap 
collapse~\cite{2023_markiewicz_arxiv_gapsend2_mott_slater_cuprate}. 
Note that the pseudogap collapse with pressure appears to be at a 
higher doping than that found in Ca-substituted 
YBCO~\cite{2001_Tallon_YBCO_pseudogap_collapse}, 
which may be due to pressure pushing the $d_{z^2}$ band above the Fermi 
level while preserving the $d_{x^2-y^2}$ band occupation, 
see SM Sec.~S5.1~\cite{SM} for discussion.

Using the prototypical stoichiometric cuprate 
superconductor YBCO$_7$, 
our study explores in-depth evolution of the electronic structure on 
a first principles basis when this cuprate is doped continuously via 
pressure to a higher Cu valency configuration with active $d_{z^2}$ 
orbitals. 
Under high pressure (100\,GPa), 
YBCO$_7$ is shown to become compatible with the second-dome SC model 
proposed in Ref.~\cite{2019_Scalapino_PRB_2nd_dome_Cu3+};
the pressure could presumably be lowered below 100\,GPa via chemical 
doping to allow access to the second dome under experimentally more 
accessible pressures.
Our analysis show that pressure leads to significant intraplanar
Cu--O hole transfer and indicates that the planar oxygen-hole content
correlates with $\Tc$ for the second dome,
as is the case for the first dome.
All studied magnetic phases are found to become unstable 
around the pressure range of the proposed second dome in YBCO$_7$,
which is a signature of pseudogap collapse,
hinting at a new connection between the second dome and the pseudogap 
phase. 

\nocite{1996_PBE_functional,
2018_Poloni_YBCO7_DFT_X-ray,
2006_Liang_YBCO_doping_formula,
1987_Ong_LSCO_Hall_effect,
2012_Zeljkovic_Oint_imaging}

\begin{acknowledgments}
We thank Yubo Zhang for his contribution in the early phase of the work 
and Sugata Chowdhury for useful discussions. 
J.~N. thanks Osk.~Huttunen foundation and INERCOM platform for financial 
support. 
The work at Northeastern University was supported by the 
U.S. Department of Energy (DOE), 
Office of Science, 
Basic Energy Sciences Grant No.~DE-SC0022216 and benefited from 
Northeastern University's Advanced Scientific Computation Center 
and the Discovery Cluster and the National Energy Research Scientific 
Computing Center through U.S. DOE Grant No.~DE-AC02-05CH11231. 
The work at Los Alamos National Laboratory was carried out under 
the auspices of the U.S. DOE National Nuclear Security Administration 
under Contract No.~89233218CNA000001. 
It was supported by the LANL LDRD Program, 
the Quantum Science Center, 
a U.S. DOE Office of Science National Quantum Information 
Science Research Center, 
and in part by the Center for Integrated Nanotechnologies, 
a U.S. DOE BES user facility, 
in partnership with the LANL Institutional Computing Program 
for computational resources. 
Additional computations were performed at the National Energy 
Research Scientific Computing Center (NERSC), 
a U.S. DOE Office of Science User Facility 
located at Lawrence Berkeley National Laboratory, 
operated under Contract No. DE-AC02-05CH11231 
using NERSC award ERCAP0020494. 
The work at Tulane University was supported by the U.S. 
Office of Naval Research Grant No.~N00014-22-1-2673. 
B.~B. acknowledges support from the COST Action CA16218. 

\end{acknowledgments}

\end{document}